\documentstyle[11pt,newpasp,twoside]{article}
\def\edcomment#1{\iffalse\marginpar{\raggedright\sl#1\/}\else\relax\fi}
\reversemarginpar

\begin{document}
\title{Simulations of Merging Clusters of Galaxies}
 \author{Sabine Schindler}
\affil{Astrophysics Research Institute, Liverpool John Moores
 University, Twelve Quays House, Birkenhead CH41 1LD, U.K.}
%\author{Ima Co-Author}
%\affil{The Name of My Institution, The Full Address of My Institution}

\begin{abstract} 
Numerical simulations of cluster mergers 
reveal many characteristics of the merging process: 
shock structure and strength, observational signatures of the
dynamical state, effects on the mass determination, turbulence and the
evolution of the X-ray luminosity and the
magnetic field.
In this article I review the results obtained from various simulations
over the last years. 
\end{abstract}

\section{Introduction}

Irregular cluster morphologies in many X-ray images of cluster as well
as indications from optical observations imply that many clusters are
not relaxed. Hence merging is a common phenomenon in clusters of galaxies. 
Such mergers of subclusters are very energetic events, which affect
clusters strongly, e.g. shocks emerge which are important because
they are the major heating
source for the intra-cluster gas and moreover they can accelerate particles.

While observations provide only snapshots of the different
evolutionary stages of mergers, simulations are the
only way to follow the evolution.
The models reveal where shocks emerge, where they move and
how strong they are. 
With the help of simulated X-ray images at different stages
of a merger a connection between X-ray morphology and the dynamical
state can be established and the effects on mass determination can
be investigated. Magnetohydrodynamic simulations show how the magnetic
field is enhanced during mergers and how the overall structure of the
magnetic field is changed due to a merger.

\section{Simulation Methods}

In order to perform realistic simulations three-dimensional
calculations are required and the different cluster
components must be taken into account. It is necessary to follow the
evolution of the dark matter as well as the intra-cluster gas. Dark
matter (and galaxies) can be regarded as collisionless particles and
can therefore be modeled by N-body simulations. The gas is best
simulated by hydrodynamic calculations. For these hydrodynamic 
calculations two different
methods have mainly been used: (1) Smoothed Particle Hydrodynamics
(Lagrangian approach, e.g. Evrard 1990; Dolag et al. 1999; Takizawa
1999; Takizawa \& Naito 2000) in which the gas is treated as particles
and (2) grid codes (Eulerian approach, e.g. Schindler \& M\"uller
1993; Bryan et al. 1994; Roettiger et al. 1996, 1997, 1998, 1999a,b;
Ricker 1998; Quilis et al. 1998) in which the simulation volume is
divided into cells. Fortunately, the choice of simulation method
seems to be irrelevant. Calculations with both methods
yield very similar results.

\section{Shocks}

Mergers produce shocks in the intra-cluster gas. These shocks are of
particular interest for particle acceleration models. The strongest shocks
emerge after the collision of subclusters, when these shock propagate
outwards along the original collision axis (Schindler \& M\"uller
1993; Roettiger et al. 1999a). However, even these shocks are
relatively mild shocks with a maximum Mach number of about 3. When
a dense subcluster falls into a cluster a shock emerges
already before the core passage: a bow shock is 
visible in front of the infalling subcluster (Roettiger et al. 1997). 
In general, the shock structure is found to be more filamentary at
early epochs and quasi-spherical at low redshifts (Quilis et al. 1998).

Observationally, the shocks are best visible in temperature maps,
because they show up as steep temperature gradients. For such maps
spatially resolved X-ray spectroscopy is necessary which can be
performed now with high accuracy with the new X-ray observatories XMM
and Chandra.

\section{Other Effects of Mergers}

Apart from shocks, mergers have many other interesting effects. For
example, the X-ray luminosity 
increases during the collision of two subclusters (Schindler \&
M\"uller 1993). The reason is that the gas is compressed, i.e. the gas
density is increased, and as the X-ray emission is proportional to the
square of the density we see enhanced X-ray emission during the passage.
During the core passage and at each rebounce an increase in the magnetic
field is visible (Dolag et al. 1999, 2000 (in prep.); Roettiger et
al. 1999b; see also Section 6).
Mergers also cause a lot of turbulence. Off-centre collisions produce
in addition angular momentum (Ricker 1998; Roettiger et al. 1998). The
shocks heat primarily the ions as has been shown in simulations which
treat ions and electrons separately (Chi\`eze et al. 1998; Takizawa 1999).

Observationally mergers cannot only be identified by multiple X-ray
maxima, but also by isophote twisting with centroid shift and
elongations: the dark matter component is always elongated along the
collision axis, while the gas is first elongated along the collision
axis but during the core passage it is pushed out perpendicular to the
collision axis, so that later an elongation perpendicular to the
collision axis can be seen (Schindler \& M\"uller 1993). Also offsets
between the collisionless component and the gas have been found
(Roettiger et al. 1997).

\section{Mass Determination in Merging Clusters}

Mass determination in clusters with the X-ray method can be affected
strongly during mergers (Evrard et al. 1996; Roettiger et al. 1996;
Schindler 1996). The reason is that during the merging process 
there can be quite strong
deviations from two assumptions necessary for the mass
determination -- hydrostatic equilibrium and spherical symmetry. For
example, at
the positions of shocks the gas is not in hydrostatic
equilibrium. Shocks cause gradients -- both in the temperature and in
the density -- and can cause therefore an overestimation of the
mass. Locally, this can lead to a mass estimate up to two times the true
mass. Substructure on the other hand tends to flatten the azimuthally
averaged profile and hence leads to an underestimation of the mass, in
extreme cases to deviations of 50\% of the true mass.

In some cases, these deviations can be corrected for, e.g. in
clusters in which substructures are well distinguishable, the
disturbed part can be excluded from the mass analysis and a good
mass estimate can be obtained. But in general, mass determinations in
non-relaxed clusters should be done very cautiously.

\section{Simulations of Mergers with Magnetic Fields}

Radio halos require the existence of magnetic fields in clusters 
on scales of a few
Mpc. Therefore magnetohydrodynamic calculations have been performed
(Dolag et al. 1999; Roettiger et al. 1999b). It has been found that
the initial field distribution is irrelevant for the final structure of
the magnetic field. The structure is dominated only by the cluster
collapse. Faraday rotation measurements can be reproduced by the
simulations for
magnetic fields of the order of $\mu$G. Most important for the
amplification of the magnetic field are 
shear flows, while the compression of the
gas is of minor importance. Mergers change the local magnetic field
strength as well as the structure of the cluster-wide field. At early
stages of the merger, filamentary structures prevail, which break
down later and leave a stochastically ordered magnetic field.

\end{document}